# Interfacial studies in CNT fibre/TiO$_2$ photoelectrodes for efficient H$_2$ production

Alicia Moya,[a] Mariam Barawi,[bz] Belén Alemán,[cz] Patrick Zeller,[d] Matteo Amati,[d] Alfonso Monreal-Bernal,[c] Luca Gregoratti,[d] Víctor A. de la Peña O'Shea,[b] and Juan J. Vilatela[c*]

[a]Dpto. Física Materia Condensada (Módulo 3), Universidad Autónoma de Madrid, 28049, Madrid, Spain.
[b]MDEA Energy Institute, Avda. Ramón de la Sagra, 3, 28937, Mostoles, Madrid, Spain.
[c]IMDEA Materials Institute, Eric Kandel, 2, 28906 Getafe, Madrid, Spain
[d]Elettra-Sincrotrone Trieste S.C.p.A., SS14-Km163.5 in Area Science Park, 34149 Trieste, Italy.
[e]juanjose.vilatela@imdea.org, victor.delapenya@imdea.org

**Abstract**
An attractive class of materials for photo(electro)chemical reactions are hybrids based on semiconducting metal oxides and nanocarbons (e.g. carbon nanotubes (CNT), graphene), where the nanocarbon acts as a highly-stable conductive scaffold onto which the nanostructured inorganic phase can be immobilised; an architecture that maximises surface area and minimises charge transport/transfer resistance. TiO$_2$/CNT photoanodes produced by atomic layer deposition on CNT fabrics are shown to be efficient for H$_2$ production (0.07µmol/min H$_2$ at 0.2V vsAg/AgCl), nearly doubling the performance of TiO$_2$ deposited on planar substrates, with 100% Faradaic efficiency. The results are rationalised based on electrochemical impedance spectroscopy measurements showing a large reduction in photoelectron transport resistance compared to control samples and a higher surface area. The low TiO$_2$/CNT interfacial charge transfer resistance (10Ω) is consistent with the presence of an interfacial Ti-O-C bond and corresponding electronic hybridisation determined by spatially-resolved Scanning Photoelectron Microscopy (SPEM) using synchrotron radiation.

# 1. Introduction

There is pressing need to develop new materials for the efficient conversion of solar energy into clean fuels, chemical or electrical energy; central to all future worldwide energy management schemes. Increasing the overall efficiency of solar conversion cycle requires materials that can effectively carry out multiple sequential processes, including: efficient absorption of light, separation of photogenerated carriers, transport of photocarriers to active sites, and performing the desired chemical reaction.

An attractive class of materials for photo(electro)chemical reactions are hybrids based on semiconducting metal oxides and nanocarbons (e.g. carbon nanotubes (CNT), graphene),[1] successfully used to improve efficiency in related systems including dye-sensitised solar cells (DSSCs),[2] electrocatalytic water splitting,[3, 4] photocatalytic hydrogen production,[5, 6] as well as in sensors,[7] batteries[8] and supercapacitor devices.[9] In general, the nanocarbon acts as a highly-stable conductive scaffold onto which the inorganic phase is immobilised. Keeping the thickness/size of the inorganic phase (<100 nm) increases surface to volume ratio of the inorganic, desired for light absorption and catalytic activity, while simultaneously reducing charge transport resistance and transfer across the large inorganic phase/nanocarbon interface. Indeed, electro chemical impedance spectroscopy measurements consistently show a low charge transport/transfer resistance in various metal oxide/nanocarbon hybrid systems [10, 11, 12]. There is, in addition, the widespread view in the literature that separation of photogenerated carriers is favoured through differences in electronic structure of the two phases in these hybrids, as experimentally observed by transient absorption spectroscopy measurements on selected nanocarbon/$TiO_2$ hybrid photo(electro)catalysts. [13, 14, 15]

In the quest towards more efficient nanocarbon-based hybrid materials for photocatalytic process, it is key to improve our understanding of how the hybrid structure affects photocatalytic activity. Issues such as the effect of electronic structure of the hybrids, the nature of the interaction at the nanocarbon/MOx interface and the hybrid surface chemistry on photocatalytic activity, are still unresolved. A suitable test system to conduct such studies are hybrids consisting of porous membranes of low-defect CNTs coated with nanostructured layer of $TiO_2$,[16, 17] which remains the archetypal photoanode. [18] In particular, large-area CNT/$TiO_2$ photoelectrodes produced by atomic layer deposition (ALD) of $TiO_2$ on pre-formed arrays of CNTs [19, 20] are ideal systems for this endeavour; they combine: a low defect density in the nanocarbons and large crystalline domain size of $TiO_2$ (in lateral direction), low charge transport resistance, and a robust free-standing format enabling photoelectrochemical studies. The two phases are strongly coupled; through advanced energy electron loss spectroscopy (EELS) and transmission electron microscopy studies on ALD $TiO_2$/CNT hybrids, Deng et al. convincingly demonstrated the presence of an interfacial Ti-O-C bond,[19] predicted to be stable (1.7eV ) according to simulations[21] and hinted experimentally in other reports. [22, 23, 5, 20] Similarly, in a previous work on ALD $TiO_2$/CNT hybrids we determined the resistance for transfer of photoexcited electrons through $TiO_2$ and into the percolated CNT current collector network to be around 10 Ω.[20] In spite of these advances, there are no studies that relate the properties of these hybrid materials to their performance as photoelectrodes for solar energy conversion. Here, we present a detailed study showing high photoelectrochemical activity of $TiO_2$/CNT photoanodes for $H_2$ production and rationalise the results based on electrochemical impedance spectroscopy measurements (EIS) and interfacial properties determined by spatially-resolved Scanning Photoelectron Microscopy (SPEM) using synchrotron radiation.

# 2. Experimental

## 2.1. Material

To produce $TiO_2$/CNTF hybrids, CNTF macroscopic film is used as substrate where $TiO_2$ is deposited by ALD process. In first place CNT fibre is synthesised by the direct spinning of carbon nanotubes[24, 25] grown by floating catalyst CVD. Butanol, ferrocene and thiophene are used as sources for carbon, catalyst (Fe) and promoter (S) respectively. The reaction is produced under hydrogen flow at 1250°C in a vertical furnace, where the mm long CNTs grow and form an elastic aerogel that is directly spun. The fibre used in this work is produced with an initial solution composition of 0.8 wt.% of ferrocene, 1.5 wt.% of thiophene and 97.7 wt.% of butanol, producing predominantly few-layer CNTs with nearly random in-plane orientation. Once the macroscopic film is produced it is subjected to several ALD cycles using tetrakis(dimethylamido)titanium(IV) from Sigma-Aldrich without any modification as an inorganic precursor. One ALD cycle consists in the injection of precursor (0.1s pulses) into the chamber at 135°C, Ar purging (35s) and oxidation treatment (20s oxygen plasma), with a growth rate of 0.05nm per cycle. The cycle is repeated up to the desired thickness (10, 20, 60, 80, 100 nm in this study). These conditions lead to the deposition of an amorphous layer of $TiO_2$ that is then annealed at 400°C for 1 h in Ar atmosphere to obtain crystalline anatase. This ALD process was carried 75 out in a Fiji F200 (Cambridge NanoTech, USA) equipment. A control sample of $TiO_2$ grown by ALD on ITO was produced under nominally identical synthesis conditions as the hybrid. Photographs of typical samples are included in supplementary information. Morphological characterisation of the hybrids was carried out using a dual beam with field-emission scanning electron microscope (FIB-FEGSEM, Helios NanoLab 600i FEI) at 5 keV.

## 2.2. SPEM

X-ray Photoelectron spectroscopy measurements were performed at ESCA microscopy beamline at Elettra synchrotron facility (Trieste, Italy) where both conventional and spatially resolved XPS were carried out using photon energy of 650.7 eV. The photoelectrons are collected with a SPECS-PHOIBOS 100 hemispherical analyser and detected by a 48-channel electron detector. In the spatially resolved operation mode the X-ray beam is focused on the sample by Fresnel Zone-Plate focusing optics and the images are acquired collecting core-level photoelectrons emitted within the relevant kinetic energy window while raster scanning the specimen with respect to the microprobe (Scanning Photo Electron Microscopy; SPEM). This mode is used to study TiO2/CNTF surface and interface with a lateral resolution of approximately 130 nm. Due to the 48-channel detector each point in the acquired maps contains a 48 point spectra spanning the Kinetic Energy window. Moreover, XPS spectra are also acquired without the beam focused as in a conventional XPS mode to obtain an average signal on an area of the sample surface of about 75 μm diameter. $TiO_2$/CNTF hybrids and pristine CNTF samples were thermally degassed in-situ (400°C, p = $10^{-9}$mbar) for several hours prior to the analyses in order to clean the surface. All C1s, Ti2p and O1s spectra were fitted by mixed Gaussian/Lorentzian peak functions and Shirley background profile was subtracted. To analyse the image the spectrum extracted from each point was fitted with the procedure described above and the intensities of each component were shown point by point in the corresponding maps. All the spectra were calibrated and represented with respect to Fermi level (binding energy of 0 eV).

## 2.3. (Photo)electrochemical measurements

Electrochemical and photoelectrochemical measurements were performed in a three-electrode glass cell with a quartz window containing an aqueous solution of 0.5 M $Na_2SO_4$ at pH=9. $TiO_2$ and $TiO_2$/CNT samples were used as working electrode. The counter electrode was a platinum wire, and the reference was an Ag/AgCl electrode. Current density (at dark and under illumination) and impedance voltage dependence were measured with a potentiostat-galvanostat PGSTAT302N with an integrated impedance module FRAII (10 mV of modulation amplitude is used at AC frequencies from 1Hz to 10000Hz). Current was normalised by projected electrode area (1cm x 1cm). During the experiment, an argon flow of 40sscm was passed through the top of the cell. A UV LED lamp (80mW/cm$^2$) was used as light source.

## 3. Results and discussion
### 3.1. Hybrid structure
The hybrid used in this work consist of large-area electrodes produced by ALD growth of a thin (10-100nm) conformal TiO$_2$ layer onto a continuous pre-synthesised CNT fibre fabric, a network that acts as both scaffold and built-in current collector. The resulting architecture maximises surface area of the metal oxide while minimising its thickness, which is beneficial to increase efficiency of surface-catalysed processes and reduce transport resistance through the metal oxide, both attractive features for photoelectrodes.

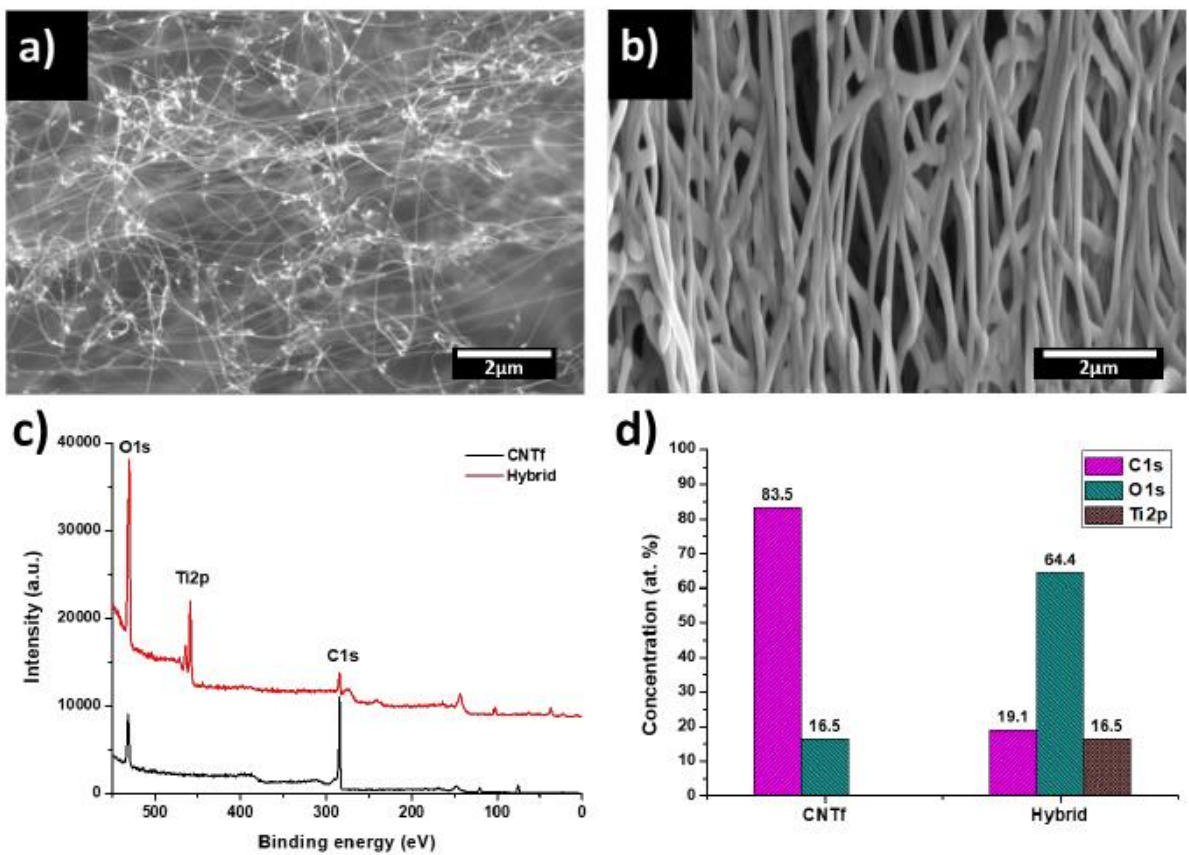

**Figure 1**: General structure of CNT fibres and CNT fibre/TiO2 hybrids. Electron micrographs showing the porous structure of a) the initial CNT fibre fabric and b) the hybrid with a 20nm conformal coating of TiO2. c) XPS survey spectra of both materials.

Fig. 1a shows the porous, yet highly-interconnected structure of the CNT network. During ALD the CNT bundles act as substrate for the nucleation of the inorganic precursor, which penetrates through the porous structure of the fibre forming a continuous TiO$_2$ coating (Fig. 1b). The growth of the TiO$_2$ layer does not disrupt the connection between CNTs, thus preserving the high longitudinal conductivity (> 10$^4$S/m) of the CNT material.

X-ray photoelectron spectroscopy (XPS) can be conveniently used to analyse the chemical composition of these hybrids. Fig. 1c shows the characteristic XPS survey spectra and surface composition of pristine CNT fibre material and an example of a hybrid with 20nm of conformal TiO$_2$ coating. The CNT fibre presents a predominant emission at about 284eV corresponding to the C1s peak from highly crystalline carbon nanotubes.[26] The hybrid also presents C1s emission along with an enhancement of the O1s peak and an additional emission at 459 eV from Ti2p states.

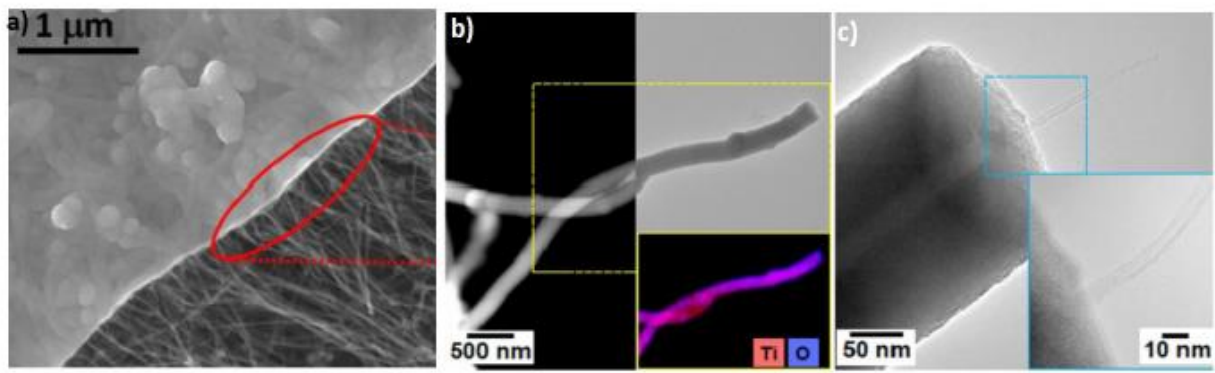

**Figure 2**: Electron micrographs of the interface between CNTs and TiO$_2$ in areas where the inorganic part of the composite structure cracked and led to CNT pull out.

It is of particular interest to characterise the interface between CNTs and TiO$_2$. From extensive observation of hybrid samples under electron microscopy, we have identified fractured areas in the material as ideal for this purpose. Fig. 2 shows a typical example of such a region in a thin hybrid sheet sample purposely fractured. The images show clearly the presence of a crack that has propagated through the metal oxide phase and led to exposure of CNTs pulled out from the TiO$_2$ coating at the opposite side of the sample. Further inspection by high-resolution transmission electron microscopy (HRTEM) shows that cracked areas consist in fact of the abrupt termination of the TiO$_2$ coating around CNTs. The images in Fig. 2b,c present an example of a CNT protruding from a bundle coated with TiO$_2$ showing clearly how these fractured areas expose the CNT/TiO$_2$ interface, a region that is otherwise very challenging to access for characterisation.

We study the CNT fibre/TiO$_2$ hybrids in areas near such cracks and thus gain insight into the chemistry at the CNT fibre/TiO$_2$ interface. Fig. 3a presents an example of a fractured region as observed by SEM that also enables XPS analysis. The region with TiO$_2$ coverage can be distinguished from the bare CNT bundles. Our strategy is to analyse such fractured regions by spatially resolved XPS spectro-microscopy. The high spatial resolution available for these measurements, 130nm, enables identification of features as small as individual CNT bundles, as shown in the C1s emission map in Fig. 3b. The chemical maps for the whole probed area, obtained by setting the energy window of the analyser to the kinetic energies of the C1s (Fig. 3c) and Ti2p emissions (Fig. 3d), confirm the transition from a TiO$_2$-rich to CNT-rich area. The transition is more prominent when taking the Ti2p/C1s intensity ratio to remove topographical effects, which leads to a better resolved chemical map for further analysis (Fig. 3e). Thus, the XPS Ti2p/C1s map confirms that the XPS chemical map has correspondence with the morphology observed by electron microscopy, particularly near transition between CNT and TiO$_2$-coated CNTs.

Once the different chemical regions are identified, C1s, O1s and Ti2p core levels spectra are acquired using two different modes: non-focused and focused X-Ray beam, referred as OSA and ZP, respectively. Having both modes provides the benefit of combining rapid screening over large areas, with high spatially resolved measurements. And very importantly, it enables continuous comparison of spectra to detect possible beam damage induced in ZP mode.

C1s emission (OSA and ZP spectra) for pristine CNTF and TiO$_2$/CNTF material are presented in Fig. 4a and their corresponding fitting in (Fig. 4b (detailed fitting parameters are shown Table S1). For pristine CNTF, both spectra are very similar, with a sharp peak at around 284.4 eV that corresponds to sp$^2$-C together with the appearance of the nanotube characteristic _π-π* loss band (290.1 eV). When analysing the spectra of TiO$_2$/CNTF material, the ZP spectra acquired at the TiO$_2$/CNTF interface (i.e. at the crack area) shows the emergence of the nanotube bundles as a sharpening of the main C1s peak from sp$^3$-C to sp$^2$-C (284.4 eV). This is distinctly different from OSA spectra corresponding to remaining adventitious C, which gives a different peak position and width of the C1s region. By taking the XPS spectra of the Ti2p a sensitivity of this material towards the X-Ray beam was found (see Fig. S1). The high density of photons in the focused X-ray beam causes a very fast (<1 min) reduction of the $TiO_2$ by forming $Ti^{3+}$ and $Ti^{2+}$ species (see SI). The strategy adopted to minimise this reduction/beam effect is to reduce the illumination time. This can be done by using the image mode of the SPEM where each point is illuminated for a short period and then the beam moves to a fresh spot. A spectrum can be extracted from each pixel of the image. Using short integration times (here 50 ms per pixel) the beam effect is minimised and the extracted spectra could be analyzed. In addition, we have performed a fitting analysis at each point of the Ti2p chemical maps to determine the distribution of the Ti oxidation states (see experimental part).

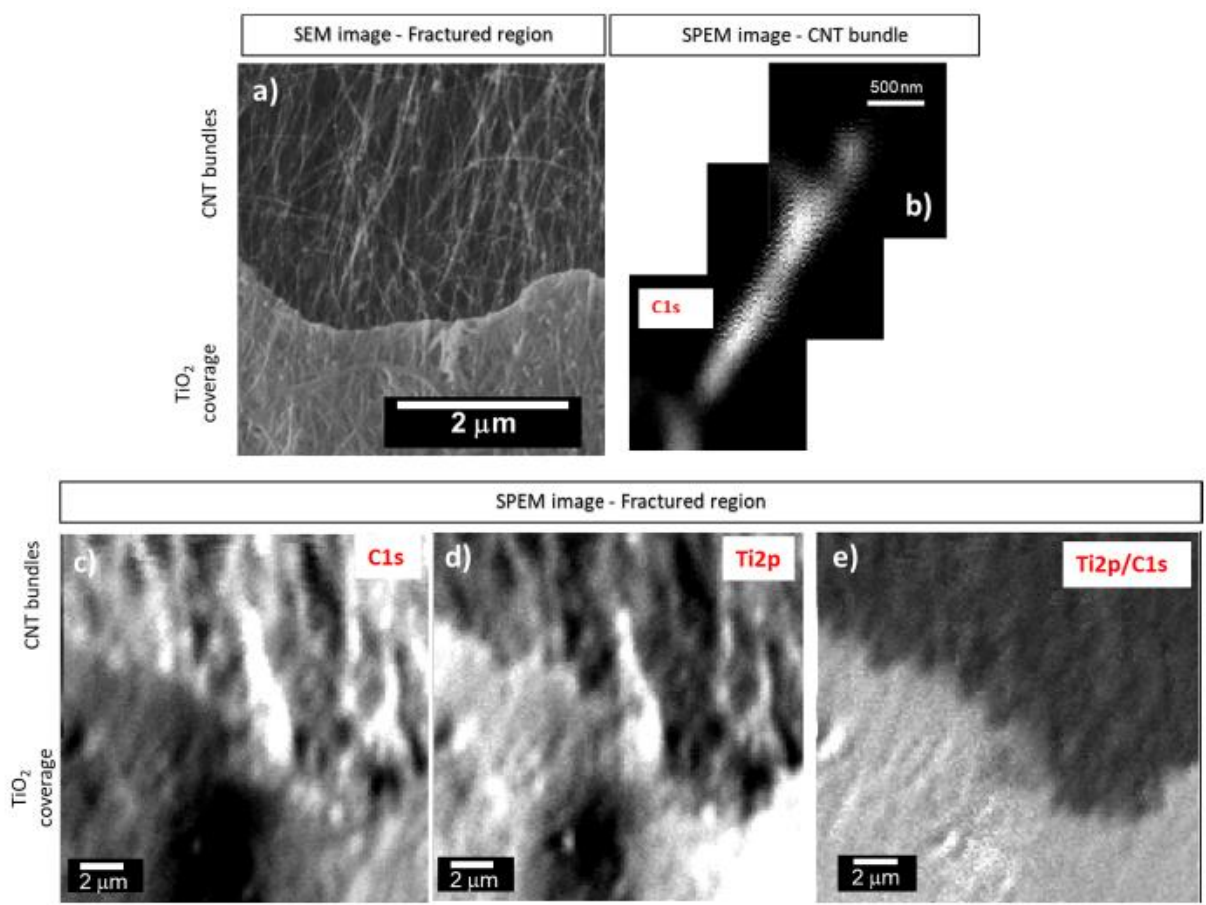

**Figure 3**: Chemical mapping of fractured regions exposing the TiO$_2$/CNT interface. a) SEM image of a representative, intentionally fractured region of the TiO$_2$/CNTF hybrid. SPEM mappings showing b) CNTF bundle observed from C1s emission, and fractured region of the TiO$_2$/CNTF hybrid mapped through c) C1s, d) Ti2p and e) Ti2p/C1s emissions.

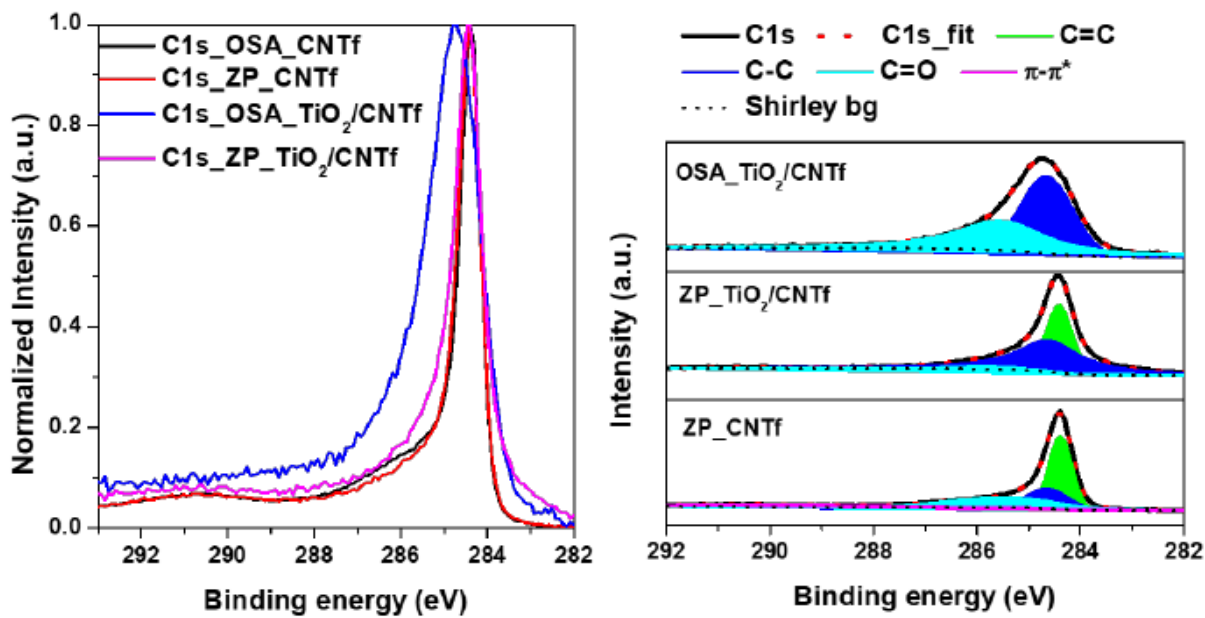

**Figure 4**: a) XPS spectra of C1s region for TiO2/CNTF hybrid and pristine CNTF using non-focused (OSA) and focused X-Ray beam (ZP) and b) corresponding fitting spectra.

Fig. 5a shows the extracted chemical map of Ti2p region of the fractured region in Fig. 3 together with their $Ti^{4+}$ (at 459 eV) and $Ti^{3+}$ (at 457.5 eV) oxidation states maps. The hybridisation between $TiO_2$ and CNT bundles can be identified by the ratio between the $Ti^{4+}$ and $Ti^{3+}$ oxidation states (Fig. 5d). The area ratio of these components increases at the border between the $TiO_2$ coverage and CNT bundles, with an enhancement at the $TiO_2$/CNTF interface. Selected regions, marked as A-B-C in Fig. 5d, were subjected to further analysis, as presented in Fig. 5e. The most relevant changes are observed for the $Ti^{3+}$ component, whose intensity increases significantly at the $TiO_2$/CNTF interface due to the formation of non-stoichiometric oxide, most probably via Ti-O-C bonds. Additionally, a new contribution related to $Ti^{2+}$ states appears in the low binding energy range (455.2 eV), but which is probably related to the reductive effect of the X-ray beam (see $Ti^{2+}$ map in Fig. S2). More chemical mapping of different $TiO_2$/CNTF hybrid samples have been analysed and are shown in Fig. S3.1 and S3.2.

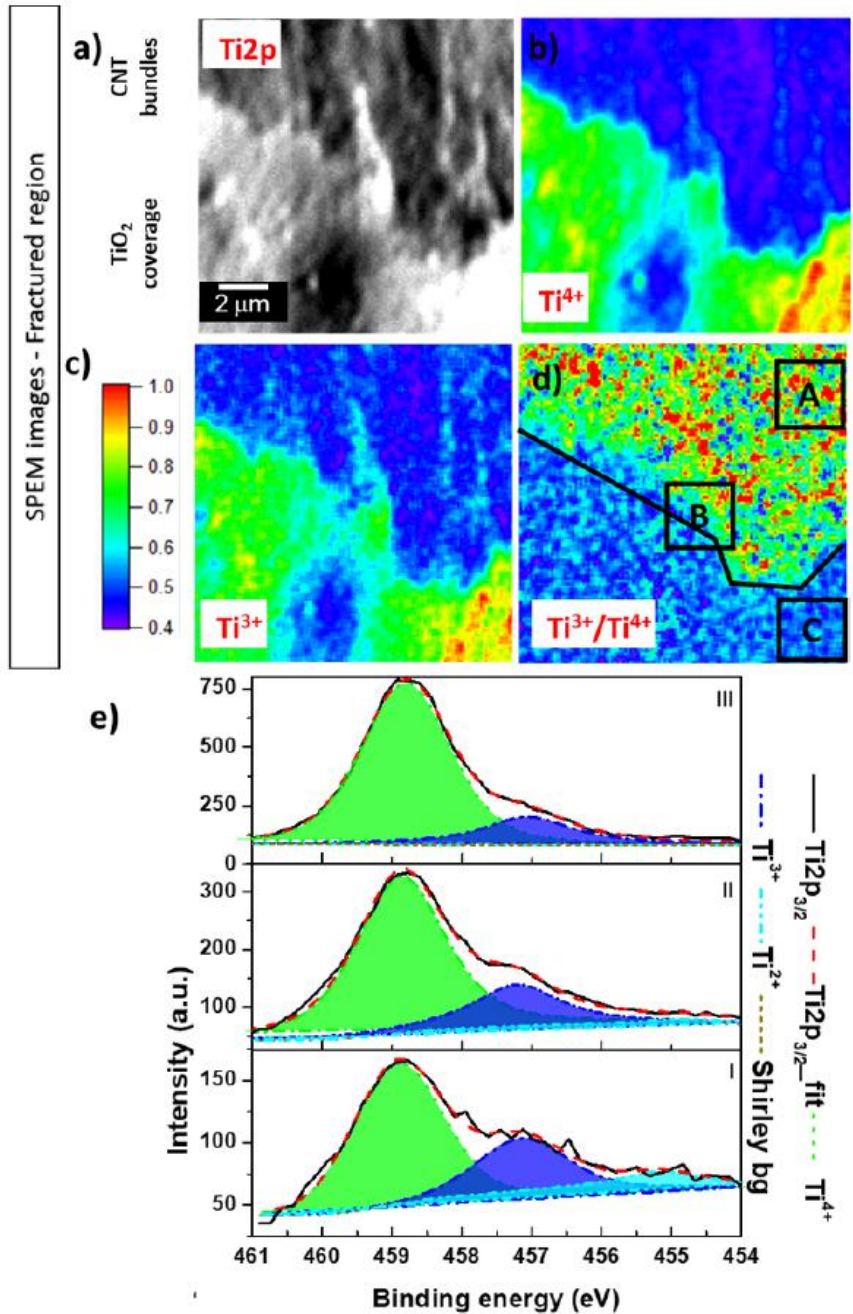

**Figure 5**: SPEM micrograph of a fractured area of TiO$_2$/CNTF showing the spatial distribution of a) Ti2p emission, b) Ti$^{4+}$ area, c) Ti$^{3+}$ area, d) Ti$^{3+}$/Ti$^{4+}$ area ratio mapping, highlighting the A-B-C regions where fit spectra in e) were extracted and labeled as I-II-III, respectively.

## 3.2. Valence band analysis

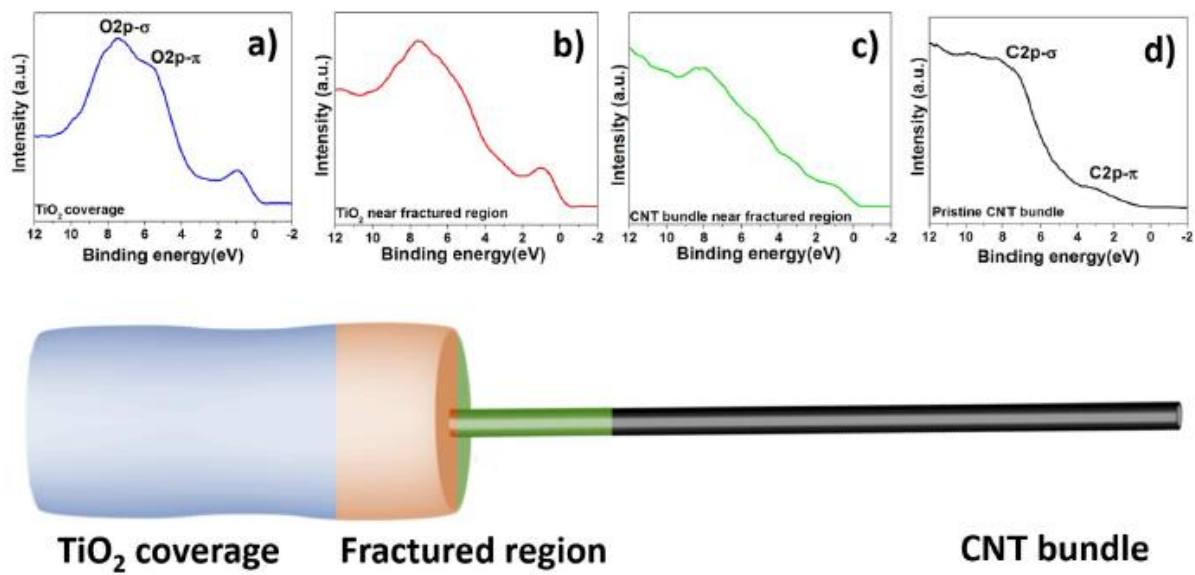

**Figure 6**: Valence band spectra when moving from the TiO$_2$-coated area, through the interface, to the CNT bundles. VB spectra for CNT fibres (d) and for TiO$_2$ on ITO (a) included for reference. The spectra at the hybrid interface shows the presence of "inter band gap" states which is consistent with the formation of a covalent Ti-O-C bond and overall, proving the strong coupling between the two components and electronic hybridisation.

It becomes of interest then to analyse the electronic structure of the hybrid, which considering the small thickness of the TiO$_2$ and CNTF phases and the presence of an interfacial Ti-O-C bond, could differ from a bulk heterojunction governed by electronic band alignment. Following the same strategy to characterise the TiO$_2$/CNT interface as discussed above, in Fig. 6 we present changes in valence band spectra when moving the probing area from the fully covered TiO$_2$/CNT hybrid, across the interface and onto the almost uncoated CNTs.

It is instructive to consider first the spectra for the separate phases. For the pristine CNT material the valence band emission (Fig. 6d) presents the features of C2p-π states and C2p-σ states at around 3 and 8 eV respectively, with a sharp increase of the density of states from Fermi level characteristic of its metallic behaviour. For TiO$_2$ its valence band spectrum (Fig. 6a) shows the expected contribution of O2p-π states and O2p-σ states at 6 and 8eV respectively, together with an onset of the valence band edge at 3eV, which agrees well with the bulk band-gap energy of TiO$_2$. At the interface, the hybrid TiO$_2$/CNT fibre material presents a valence band spectra with mixed features from the individual phases as well as the emergence of new features. O2p-σ states, for example, are substantially less intense at the hybrid TiO$_2$/CNT interface than in the reference TiO$_2$ (Fig. 6a), i.e. TiO$_2$ on ITO. Similarly, the VB spectra of the hybrid consistently show the presence of "inter band gap" energy states below 3eV, which are most intense at the TiO$_2$/CNT interface. These energy state match contributions from C2p-π (≈3 eV) and from Ti3d (≈1 eV) states from reduced titania. Both contributions are consistent in our material through formation of a Ti-O-C bond during hybridisation together with the contribution from beam damage (see such effect in Fig. S4.1). A comparison of the valence band line-shape of the hybrid samples with that a sample of TiO$_2$ with extensive beam damage provides a qualitative method to discriminate damage from hybridisation effects (Fig. S4.2). After such analysis, we can identify changes in Ti3d states and in the C2p-π peak along the $TiO_2$/CNT interface (Fig. 6b,d) as effective indication of hybridisation.

Finally, we have simulated the VB spectra of the hybrid by combining the normalised spectra of both individual components and the results shows a very different spectra in comparison with that obtained for the hybrid material (Fig. S4.3). In particular, the main differences between the measured and simulated hybrid spectra is the increase of density of the states below the $TiO_2$ valence band, which is ascribed to the hybridisation process between $TiO_2$ and CNTF. Overall, these results confirm that the $TiO_2$/CNT material is a hybrid with strong coupling of the two phases and the development of electronic structure features distinct from those of $TiO_2$ or CNT alone. Next, we study the extent to which these hybrid architecture translates into more efficient photoelectrochemical $H_2$ production through improved charge separation/transport when used as a photoanode.

## 3.3. (Photo)electrochemical measurements

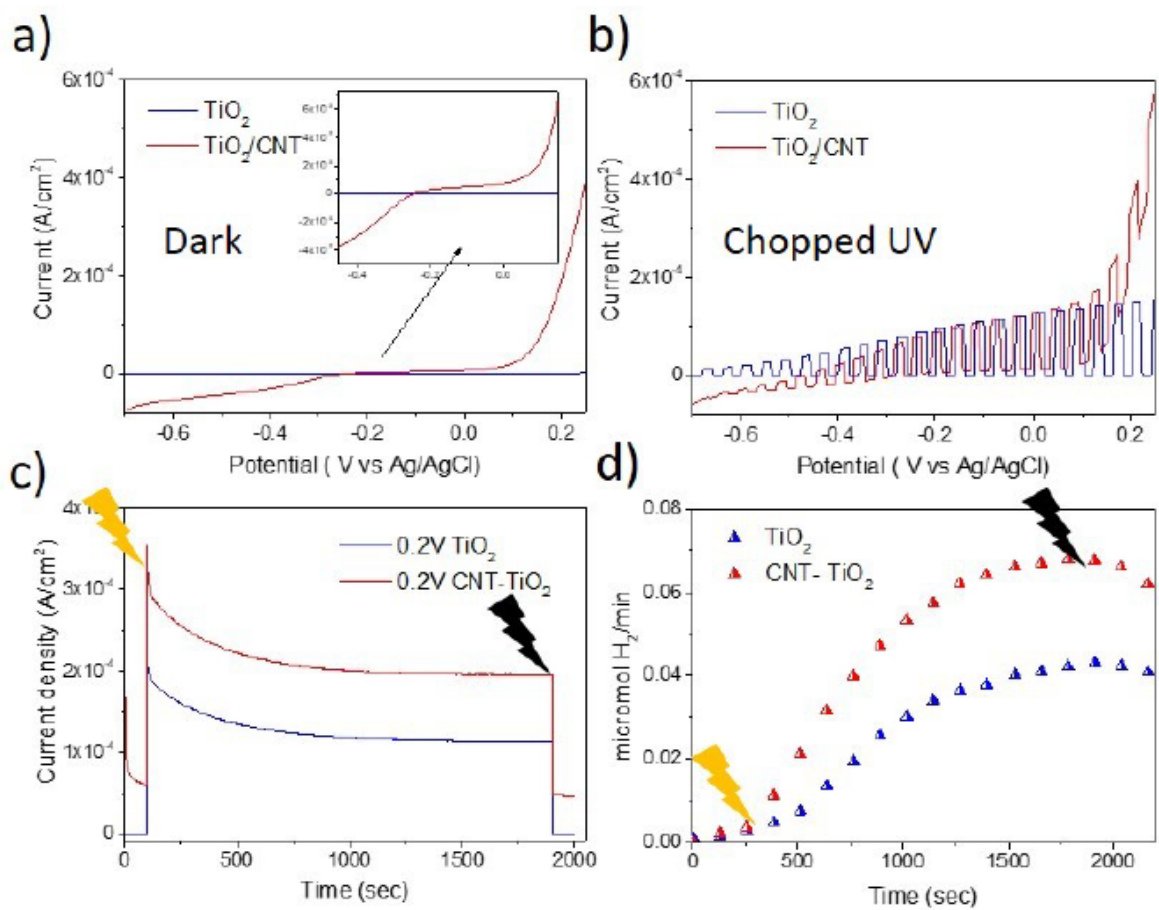

**Figure 7**: a) Linear sweep voltammetry of $TiO_2$ (blue) and CNT-$TiO_2$ Hybrid (red) photoelectrodes under a) dark conditions and b) chopped UV illumination at 10 mV/sec. c) Current densities achieve under 0.2V and UV illuminations during c) Hydrogen Evolution Reaction (HER).

Photoelectrochemical measures in aqueous solution of 0.5M $Na_2SO_4$ at pH = 9 in dark and under UV irradiation were performed on both the CNTF/$TiO_2$ hybrid and the control sample of $TiO_2$ on ITO (Photograph in Fig. S5). Fig. 7a,b show several linear voltammograms (LSV) performed at 10 mV/sec to explore the currents and photocurrents densities at different bias potentials. As shown in Fig. 7a, hybrid samples show much higher current densities at lower potentials. This effect is observed at all different scan rates used for linear sweep voltammetry measurements from 1mV/s to 100mV/s (Figure S10). We attribute the higher current density in hybrids to their substantially higher electrical conductivity and higher capacitance compared to the thin-film control sample. The onset for oxidation and reduction under dark conditions shows a much lower overpotential compared to the control samples, leading to current densities around 20 percent higher in the CNTF/$TiO_2$ hybrid than in $TiO_2$ on ITO.

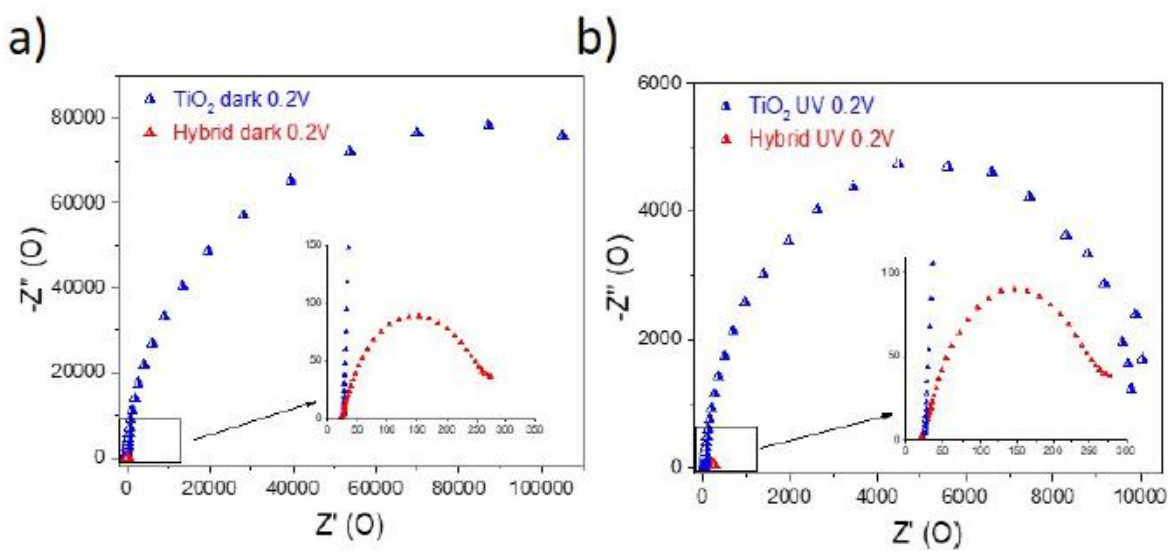

**Figure 8**: Nyquist plots obtained by Electrochemical Impedance Spectroscopy (EIS) of TiO2/ITO (red one) and TiO2/ITO/CNT (blue one) samples at 0.2 V vs Ag/AgCl.

The results obtained under chopped UV irradiation show a similar efficiency in terms of light absorption for both materials. As observed in Fig. 7b the detected photo-response near equilibrium potential is not significantly affected by the presence of CNTs in the hybrid. Note that for this sample, with a $TiO_2$ thickness of 100nm, the volume fraction of $TiO_2$ is above 50%, which implies that even if CNTs are strong light absorbers most UV light is harvested by the semiconducting metal oxide. Consistent with LSV measurements, at larger positive and negative potentials there is a pronounced increase in current density, as also observed under chronoamperometric conditions (Fig. S6). This behaviour demonstrates that the CNT fibre/$TiO_2$ hybrid presents improvements in both the electrical transport and charge separation.

Next, the $TiO_2$/CNT hybrid photoelectrodes were evaluated for hydrogen evolution reaction (HER) at different bias potentials under UV illumination by directly measuring $H_2$ through a gas chromatographer. HER at 0V is presented in Fig. S7. In that case the HER is also higher in the hybrid case, but the difference is lower than the one obtained at 0.2V, as expected at these both potentials. In fact, Fig. 7c,d, show this increase in terms of current density which translates into a higher hydrogen production during the reaction of 0.07µmol $H_2$/min compared with 0.04 for the benchmark sample. Taking into account the current densities obtained during the reaction and the quantity of evolved hydrogen, Faradaic efficiency is close to 100%, demonstrating that there are not other reactions taking place in the electrode. Indeed, we observed no degradation of the materials after photoelectrochemical tests (Fig S11). The increased $H_2$ production of CNTF/$TiO_2$ photoanodes is entirely consistent with the extensive characterisation of the CNT/$TiO_2$ hybrids indicating a strong interaction between the two phases and a large interface between them. We further rationalize these results by EIS measurements on the material at 0.2V vs Ag/AgCl under dark and illumination conditions (Fig. S8 and Table S2). The Nyquist plots for the hybrid show indeed a dramatic reduction in charge transfer resistance as the source of a much higher current density and higher hydrogen evolution reaction compared to the control samples of $TiO_2$, an effect also observed in related graphene/$TiO_2$ systems [27, 28]. Furthermore, a detailed look at the shape of the Nyquist plots shows the emergence of a second interface (See equivalent electric circuit in Fig. S9), with an extremely small associated resistance (lower than 10Ω). We attribute it to the $TiO_2$-CNT interface, where the two phases are covalently bonded and thus develop a hybrid electronic structure at the interface, a feature expected to favour charge separation of photogenerated carriers. Indeed, in previous work the electronic structure of the $TiO_2$/CNTf hybrid determined from UPS measurements indicated a shift of the valence band edge towards the Fermi level, although with small changes in the Fermi level itself[20].

Coupled with more efficient electron transport in the hybrid is a larger surface area of crystalline $TiO_2$ relative to the planar sample, resulting in an increase in capacitance also expected to contribute to a higher photocurrent density and $H_2$ production. The capacitance of high-surface area mesoporous nanocrystalline $TiO_2$ electrodes is often characterised by a strong dependance on electrochemical potential arising from a non-negligible chemical capacitance due to energy states (i.e. defects) between the band gap. But they are generally concentrated as an exponential distribution below the conduction band, therefore total capacitance in such systems has a much weaker potential dependence from around 0 V vs Ag/AgCl[29]. Moreover, there is a much lower defect density in the $TiO_2$ layer produced by ALD in this work, which is monocrystalline over hundreds of nanometres [20], compared to mesoporous nanocrystalline electrodes based on sintered $TiO_2$ nanoparticles. Thus, we expect that in the potential range of interest in this work (0 to 0.3V vs Ag/AgCl) chemical capacitance is not dominant.

Overall, these results highlight the effective role of CNT fibres as built-in current collectors for photoelectrochemical systems. Future dedicated photoelectrochemical studies are required to determine the extent to which electronic structure and the confluence of a higher capacitance and interfacial $TiO_2$/CNTf bonding contribute to the various charge transfer and diffusion steps involved from charge generation at the electrolyte/$TiO_2$ surface to its collection at the CNT network.

## *4. Conclusions*

This work presents a detailed study on the structure and photoelectrochemical properties of hybrid materials based on thin, highly-crystalline semiconducting metal oxides supported on highly conducting pre-formed networks of CNTs. Such architecture simultaneously maximises surface area of the active semiconducting metal oxide phase and minimises resistance by reducing diffusion length of photocarrier through. Key in this approach is the very large CNT/$MO_x$ interface. Through advanced spatially-resolved XPS using synchrotron radiation we obtain chemical maps of $TiO_2$/CNT fibre hybrids. Results show that the metal oxide is partially reduced in the hybrid, particularly at the $TiO_2$/CNT interface, consistent with the formation of an interfacial Ti-O-C bond. Analysis of spatially-resolved valence band spectra confirm that the $TiO_2$/CNT material is a true hybrid, in the sense that there is strong coupling of the two phases and the development of electronic structure features distinct from those of $TiO_2$ or CNT alone, most notably inter band gap electronic states with respect to pure $TiO_2$ grown on ITO. Used as photoanodes under UV irradiation, the hybrid material can produce 0:07µmol $H_2$/min, double that of the control sample. This is a consequence of a higher surface area of crystalline $TiO_2$, a lower over-potential and a large decrease in charge transfer resistance of photogenerated carriers. Resistance at the $TiO_2$/CNT interface, most likely assisted by the Ti-O-C bond, is vanishingly small at < 10 Ohm. An interesting perspective is to extend this hybridisation strategy to nanocarbons doped with substitutional atoms or supporting metallic nanostructured catalysts, two routes for catalytic hydrogen production which have shown promising results in related systems [30, 31, 32].

## *Conflicts of interest*
There are no conicts to declare.

## *Acknowledgements*

The authors acknowledge "Comunidad de Madrid" and European Structural Funds for their financial support to FotoArt-CM project (S2018/NMT-4367) and the Spanish Ministry of Science, Innovation and Universities through the projects Ra-Phuel (ENE2016-79608-C2-1-R), SOLPAC (ENE2017-89170-R) and FOTOFUEL (ENE2016-82025-REDT). Mariam Barawi also thanks the Juan de la Cierva Formacin Program (FJCI-2016-30567). J. Vilatela is grateful for generous financial support provided by the European Union Seventh Framework Program under grant agreement 678565 (ERC-STEM). Assistance with ALD synthesis was kindly provided by M. R. Osorio and D. Granados from IMDEA Nanoscience, and Raman characterisation kindly performed by M. Rana.

## *Appendix A. Supplementary data*
Supplementary data associated with this article can be found in the online version at http://dx.doi.org/10-XXXX/j.apcatb.XXX